\documentclass{ws-procs9x6}

\begin{document}

\title{Dark Matter}

\author{V. ZACEK}

\address{Groupe de physique des particules, \\
Universit\'e de Montr\'eal, \\ 
C. P. 6128, Montr\'eal, P.Q., Canada\\ 
E-mail: zacekv@lps.umontreal.ca}

\maketitle

\abstracts{The nature of the main constituents of the mass of the universe is one of the outstanding riddles of cosmology and astro-particle physics. Current models explaining the evolution of the universe, and measurements of the various components of its mass, all have in common that an appreciable contribution to that mass is non-luminous and non-baryonic, and that a large fraction of this so-called dark matter must be in the form of non-relativistic massive particles (Cold Dark Matter: CDM).  In the spirit of the Lake Louise Winter Institute Lectures we take a look at the latest astronomical discoveries and report on the status of direct and indirect Dark Matter searches.}

\section{Dark Matter and Astronomical Evidence}

It was in 1933 that the Swiss astronomer Fritz Zwicky working at Caltec and Mt. Palomar applied for the first time the virial theorem to eight galaxies of the Coma Cluster in order to infer the cluster's mass\footnote{Virial theorem : 2 $<E_{kin}>$+$<E_{pot}>$=0}\cite{ZW}. What he found to his surprise were peculiar velocities much larger than expected, which led Zwicky to the conclusion that apparently the Coma cluster contains at least 200 times more mass than is visible in the form of the luminous galaxies. He first coined the term "Dark Matter", which from then on became part of the vocabulary of modern cosmology. However certain traits of Zwicky's personality, who was considered by his contemporaries as brilliant, but also a bit eccentric, might be responsible for the fact that it took another 40 years until the necessity of a large quantity of dark or "hidden" mass of galaxies was accepted as a serious possibility.\\

The breakthrough occurred in the early nineteen seventies, when Vera Rubin, an astronomer at the Carnegie Institution, Washington, delivered the first clear observational evidence for dark matter as a general feature of the matter composition of galaxies. Rubin studied orbital velocities of interstellar matter in galaxies like M33 (Andromeda) and over 60 other galaxies as a function of distance to the galactic centre by measuring the Doppler shift of H$_\alpha$ - emission lines. She found that these orbital velocities did not decrease with increasing distance, but rather remained constant or even increased a little bit. This contradicts the decrease expected for a Keplerian motion, which would occur if all the mass would be concentrated in the galactic bulge region, since with
\begin{equation} 
\frac{mv_{rot}^2}{r}~=~G\frac{mM}{r^2}			
\label{eq:eq1}
\end{equation}

we would get $v_{rot}~\propto ~r^{-\frac{1}{2}}$  in this case. However, if $v_{rot}$  remains constant, Eq.~(\ref{eq:eq1}) implies that $M$ is no longer constant:  in fact, $M(r)$ must increase linearly with distance from the galactic centre ! Therefore the hypothesis was made that the galaxies were surrounded by a spherical halo of some invisible stuff, with a mass of more than 10 times the mass accounted for by luminous or gaseous matter. Amazingly for large $r$, in some cases, the halos of neighbouring galaxies seem even to overlap. \\

If we look at our own galaxy, the Milky Way, then we observe a steeply rising velocity distribution within the first 3 kLy up to 230 km/sec and then a constant or even slight increase in orbital velocity, which translates into a linear increase in mass distribution up to about 600 kLy.  The solar system is situated in the flat part of the rotation curve at a distance of about 24 kLy from the galactic centre and the Dark Matter density at this halo location is around 0.3 m$_p$/cm$^3$ (where m$_p$ is the mass of the proton, i.e. 0.94 GeV/c$^2$).  All together only 5 to 10\% of the gravitating matter appears to be in visible form in our Milky Way!\\

The Milky Way is part of the Local Group, a group of galaxies which is dominated by our Galaxy and M33, the Andromeda galaxy. The size of this group, which includes also the two Magellanic clouds and other small galaxies, is about 2.2MLy. The Milky Way and Andromeda are approaching each other with a speed of 3x10$^5$ km/h, which is explained by a gravitational pull due to the presence of at least ten times the mass of our galaxy. Moreover, the Local Group, itself located at the fringe of the Virgo Cluster (50MLy), is falling towards the latter with a speed of 1.6x10$^6$ km/h due to the gravitational pull of ten times of all visible matter! Finally the Virgo cluster itself is speeding (contrary to the Hubble flow) with 2x10$^6$km/h towards an invisible mass concentration equivalent to one million galaxies spread over 100 MLy, called the Great Attractor.\\

In the Virgo cluster itself a galaxy has been discovered in 2005 (VIRGOHI21), which contains no stars and luminous matter at all! It reveals its presence only by an HI radio frequency emission at 21cm and in this frequency window this object shows up as a hydrogen cloud with a mass of 10\% of the Milky Way. The rotation curve of this hydrogen cloud indicates a ratio of dark matter to ordinary matter (hydrogen) of at least\cite{VI} 500 to 1.\\   

Further out, at around one BLy, we find more evidence for dark matter around galaxies and galaxy clusters. The Hubble space telescope revealed that Abell 2029 is a cluster of thousands of galaxies.  Images by the Chandra X-ray satellite show that this extremely rich cluster is surrounded by a gigantic cloud of hot gas at a temperature of 10$^6$K. But in order to keep such a hot gas confined it needs at least ten times more mass than is visible! \\

\begin{figure}[ht]
  \centerline{\epsfxsize=4.1in\epsfbox{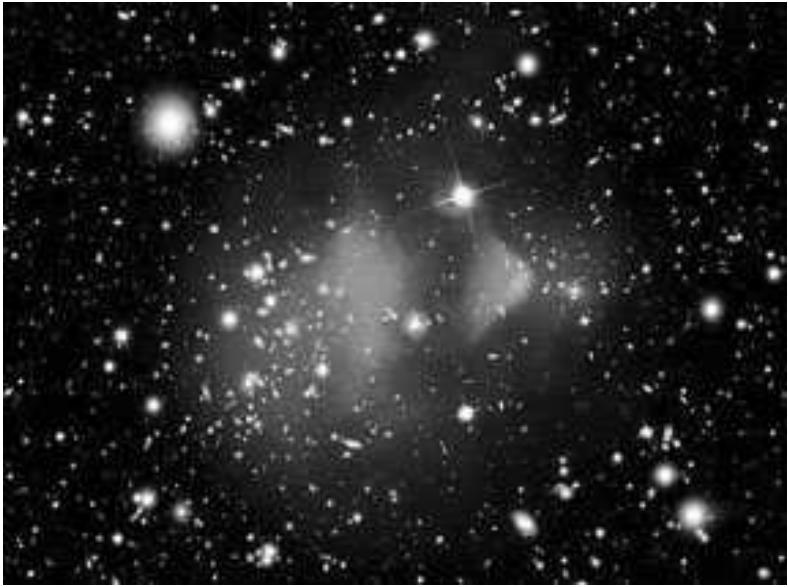}}   
  \caption{The bullet cluster is a system of two galaxy clusters, which have penetrated each other with high velocity. In the middle of the picture we see the diffuse x-ray emissions of the hot intergalactic medium created in the shock front during the collision. The distribution  of dark matter was reconstructed by gravitational lensing and coincides with the distribution of galaxies in the two cluster to the left and right of the shock front.\protect\cite{BU} \label{fig1}}
\end{figure}

Weak gravitational lensing has become another powerful tool to trace the presence of dark matter lying on the line of sight between far away light sources like quasars or bright galaxies and observers on the earth. This technique has now been developed to such a degree of perfection that it is possible to reconstruct 3D images of dark matter concentrations. The most famous example is the Bullet cluster (IE0667-56), which is about 3.4 BLy away and it is shown in fig \ref{fig1}. \\

It is a system of two clusters of galaxies, which merged through each other with a very high relative velocity of about 4.5x10$^3$km/s. During this encounter the intergalactic gas was heated in a shock front to 10$^6$ degrees Kelvin. The resulting X-ray emission was recorded in 2006 by the Chandra satellite and shows that the gas lags behind the cluster galaxies, which passed each other without collisions. Moreover gravitational lensing allowed to reconstruct the two spherical dark mater halos coinciding with the galaxies in both clusters, with the dark matter halos contributing 49 times more mass than was found in the galaxies and the surrounding gas\cite{BU}.\\

\section{Matter, Dark Energy and the Dynamics of the Universe}

Let us consider for a moment a toy universe, consisting of a large spherical region which we cut out of the real universe and which contains a large number of galaxies. Let the radius of this toy universe be $R$ and the total Mass of the sphere $M$.  Applying Gauss's law, a galaxy of mass $m$ placed at the border of our universe will have a potential energy determined by the total mass included in the sphere and concentrated at its very centre:
\begin{equation} 
E_{pot}~=~-G\frac{mM}{R}			
\label{eq:eq2}
\end{equation}

In case the vacuum is not empty, it might carry an energy density $\rho_\nu$ (due to quantum fluctuations, scalar fields or other reasons, which we do not want to discuss here). So we have to add this contribution to the potential energy. That density of "Dark Energy" can be related to Einstein's famous cosmological constant by $\rho_\nu$ = c$^2\Lambda$/8$\pi$ and we obtain
\begin{equation} 
E_{p}^\nu~=~-G\frac{m}{R}\left( \rho_\nu \frac{4}{3}\pi R^3 \right) ~=~ -\frac{1}{6}m\Lambda c^2 R^2			
\label{eq:eq3}
\end{equation}

In addition, if we allow our toy universe to expand or contract, the total energy of the border galaxy will become the sum of the kinetic and potential energies
\begin{equation} 
\frac{m}{2} \dot{R}^2 ~+~ \left(~-G\frac{Mm}{R}-\frac{1}{6} m\Lambda c^2 R^2 ~ \right)  ~=~ E_{tot}			
\label{eq:eq4}
\end{equation}

where the interesting feature of (\ref{eq:eq4}) is that $\Lambda$ adds to gravitation! What can we say about $E_{tot}$ on the right hand side? Here we have to specify the geometry of space of our toy universe and solve Einstein's equation\footnote{$R_{\mu\nu}-\frac{1}{2}g_{\mu\nu}R~=~-\frac{8\pi}{c^2}T_{\mu\nu}+\Lambda g_{\mu\nu}$}, which does precisely this: it relates mass- energy to the curvature of space. To make a long story short the result is simply
\begin{equation} 
E_{tot} ~=~ -k\frac{mc^2}{2}
\label{eq:eq5}
\end{equation}

with $k$ = 0 for a flat universe with an Euclidean geometry, $k$ = +1, for an universe with a positive curvature like on a sphere, $k$ = -1 for a hyperbolical geometry, like on a saddle.\\

Let us stop for a second and ask which forces act on the border galaxy! Since force equals minus the gradient of the potential we get
\begin{equation} 
\vec{F} ~=~ -\frac{dE_{pot}}{dR}\hat{r} ~=~ -G\frac{Mm}{R^2}\hat{r} ~+~ \frac{1}{3} m\Lambda c^2 R \hat{r}
\label{eq:eq6}
\end{equation}

and we see that the matter distribution inside our universe gives the usual attractive gravitational force (minus sign!). However, in the second term, a positive $\Lambda$ or positive "Dark Energy" density $\rho_\nu$ has the effect of a net repulsive force (positive sign), which increases with increasing dimensions of the universe! We also see that from a certain radius on, the term containing $\Lambda$ will dominate eq. (\ref{eq:eq4}).  In fact setting $k$ = 0 we get
\begin{equation} 
\dot{R}(t) ~=~ \sqrt{\frac{\Lambda c^2}{3}}R 
\label{eq:eq7}
\end{equation}

and the distance of our border galaxy will grow exponentially.\\

In order to complete the description of the dynamics of our toy universe we have to take into account still the Hubble expansion, which relates the recessional velocity $v$ of two cosmological objects to their relative distance $D$, with $v$ = $HD$. The Hubble parameter $H$ varies with time but is the same everywhere in the universe at a given time counted from the Big Bang. Today's measured value for the Hubble parameter is $H_0$ = 71 km s$^{-1}$Mpc$^{-1}$.  If we introduce Hubble's law in (\ref{eq:eq4}) by writting
\begin{equation} 
\dot{R} ~=~ H(t)~R(t) 
\label{eq:eq8}
\end{equation}

and divide through by $m$ we obtain the Friedmann equation. If we give $R$ the meaning of the radius of the universe (more precisely it is its scale parameter), this equation describes the dynamics of the real universe:
\begin{equation} 
\left( \frac{\dot{R}}{R} \right)^2 ~=~ H(t)^2 ~=~ \left[ \frac{8\pi}{3}G(\rho_m+\rho_\nu)-\frac{kc^2}{R^2} \right] 
\label{eq:eq9}
\end{equation}

where we have replaced the matter content $M$ by 4/3$\pi R^3\rho_m$. Again, since the matter density $\rho_m$ will decrease in an expanding universe, $\rho_\nu$ will take over at a certain moment and lead to the exponential expansion as described by eq. (\ref{eq:eq7}). \\

The density in the Friedman equation, which corresponds to a flat, Euclidian universe with $k$ = 0 has been dubbed the "critical" density
\begin{equation} 
\rho_c(t) ~=~ \frac{3 H(t)^2}{8\pi G}
\label{eq:eq10}
\end{equation}

\begin{figure}[ht]
\centerline{\epsfxsize=3.0in\epsfbox{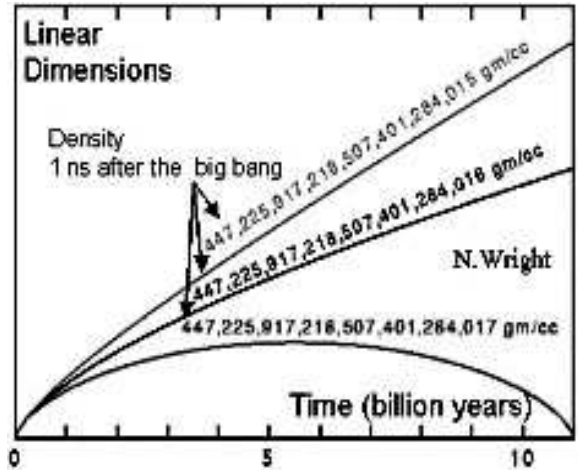}}   
\caption{Evolution of the universe as a function of the energy-matter density one nano-second after the Big Bang. The curve in the middle corresponds to the critical density in a flat, euklidian universe. Tiniest deviations from the critical density lead to drastically different evolutions of the universe. \label{fig2}}
\end{figure}

Today the critical density for $H_0$ =71 km s$^{-1}$Mpc$^{-1}$ is = 9x10$^{-27}$ kg/m$^3$ or around 6 hydrogen atoms per cubic meter. How "critical" the critical density really is for the evolution of the universe, can be read from fig. \ref{fig2}. Here the expansion of the universe is shown as a function of the total density fixed one nanosecond after the big bang. Three scenarios are given: $\rho$ precisely equal to $\rho_c$, a value of $\rho$, being 1 part in 10$^{24}$ larger and a value being 1 part in 10$^{24}$ smaller than $\rho_c$. Interestingly, in each case the history of the universe differs substantially! Thus, a tiny deviation at this very early moment results in a universe, which either would already have collapsed today or flown pretty much apart! Since such a fine tuning is required to find ourselves 13.7 Gyr after the Big Bang in the universe we live in, one might be tempted to suspect that there is a mechanism which has set the density is precisely $\rho_c$! This is the "flatness" problem in Big Bang cosmology.\\

There is also another reason to suspect that the density of the universe is exactly $\rho_c$. Wherever we look, the universe appears amazingly homogeneous and isotropic, in agreement with the "cosmological principle" which states, that at a given cosmic age the universe looks the same for all observers, wherever they are.  But in order to look the same, all corners of the universe had to be in thermal equilibrium once and therefore causally connected. But here we have a problem: a galaxy 10 GLy  to the east of us and a galaxy 10 GLy years to the west could not have exchanged a light signal in a universe which is only 13.7 GLr old! This is the "horizon" problem in Big Bang cosmology\\

The cosmic inflation scenario solves elegantly both, the "horizon" and the "flatness" problem. It posits that 10$^{-34}$sec after the Big Bang a dramatic exponential expansion of space occurred, which increased the size of the universe by a factor of 10$^{50}$. The enormous stretching removed all irregularities and flattened out all curvature. This period of inflation lasted probably just 10$^{-32}$ sec and after that the universe continued its much less exciting Hubble expansion, but tuned to the critical density\cite{CI} $\rho_c$.\\

Since $\rho_c$ plays such an important role, the quantities on the right side of eq. \ref{eq:eq9} are normalized to the critical density and the dimensionless quantities $\Omega_m$ = $\rho_m$/$\rho_c$, $\Omega_\Lambda$ = $\rho_\nu$/$\rho_c$, $\Omega_k$ = $kc^2$/$R^2$ characterize the matter density, the density of dark energy and the curvature $\Omega_k$ of the universe. They obey the simple relation  
\begin{equation} 
\Omega_m ~+~ \Omega_\Lambda ~+~ \Omega_k ~=~ 1
\label{eq:eq11}
\end{equation}

Usually one defines the sum of matter and dark energy densities separately as $\Omega_{tot}$ = $\Omega_m$ + $\Omega_\Lambda$. Then  $\Omega_{tot}$ = 1 describes a flat, Euclidian universe (with $\Omega_k$ = 0!), $\Omega_{tot}$ $>$ 1 a closed universe with positive curvature and $\Omega_{tot}$ $<$ 1 an open universe, for ever expanding and with negative curvature. A non-vanishing cosmological constant has therefore two effects: in its meaning as a vacuum energy it enhances the potential energy of gravity and leads to a smaller critical matter density; as a kind of "negative pressure" it accelerates the expansion of the universe.\\

\section{How to determine $\Omega_{tot}$ ?}

First of all we should try to count all the matter in the universe! It turns out that all matter in luminous form, like stars and shining gas, adds up to $\Omega_{lum}$ $\approx$ 0.01. This is not much, but on the other hand, given that a tiny deviation from the critical density one nsec after the Big Bang grows into an extremely large (or small) matter density today, this value of  $\Omega_{lum}~\approx$~0.01 is in fact extremely close to one! \\

The Big Bang nucleosynthesis together with data on the abundance of the light elements from deuterium to lithium, predicts accurately and confidently a contribution of baryonic matter with $\Omega_b$ = 0.044 $\pm$ 0.007!\cite{BM}. Moreover, matter appearing as dark matter in galactic rotation curves, dark matter deduced from the dynamics of clusters and the presence of hot gas in clusters, as well as dark matter inferred from gravitational lensing and galaxy flows, all add up to a total of $\Omega_m$ = 0.2 - 0.3. Therefore most of the matter seems to be a non-baryonic, exotic form of "dark matter".\\

The contribution of dark energy to $\Omega_{tot}$ has been found by studying the apparent magnitude of type 1a supernovae as a function of their red shift, i.e. distance. Since these supernovae are very bright and their absolute brightness is believed well known, they can be used as standard candles to explore large distances. Between 1998 and 2003 two groups, the Supernova Cosmological Project\cite{SCP} and the High-z Supernova Search Team\cite{HSS} have studied the brightness of far away supernovae as a function of redshift and found that the farthest objects were slightly dimmer than expected. These measurements indicate that the acceleration of the universe is not decreasing as expected from the decelerating gravitational force in a matter-dominated universe, but rather is accelerating as described by the second term in eq. \ref{eq:eq6}. The best match to the data was found for $\Omega_\Lambda$ = 0.7 and $\Omega_m$ = 0.3, which would bring $\Omega_{tot}$ to 1!\\

An alternative way to find $\Omega_{tot}$ would be to try to measure the curvature or geometry of the universe! Precisely this has been investigated by studying the anisotropy of the cosmic microwave background radiation, abbreviated CMB\cite{WMAP}. This relic radiation fills the entire universe and was emitted 300 ky after the Big Bang. At that moment photons could not ionize anymore and fell out of thermal equilibrium. Atoms became neutral and the universe transparent. The spectrum of the photons was that of a blackbody with a temperature of 6000 K. Looking today at the CMB photons, we look back straight towards the "surface of last scattering", when the photons interacted last time with matter, only that these photons appear today red shifted by the Hubble expansion to a blackbody temperature of a mere 2.7 K, contributing to the flicker noise on your TV screen\footnote{About 1\% of the noise on a TV screen if tuned to a non-transmitting station is CMB}.  \\

The temperature of these relic photons is amazingly uniform in whatever direction we look in the sky. Only at the level of one part in 10$^5$ we find slight deviations and local anisotropies, which range from patches in the sky with an angular size of a fraction of a degree to about 1$^\circ$ (the angular diameter of the moon is 0.5$^\circ$). These slight irregularities constitute a prolific source of information on cosmological parameters.\\

\begin{figure}[ht]
\centerline{\epsfxsize=3.0in\epsfbox{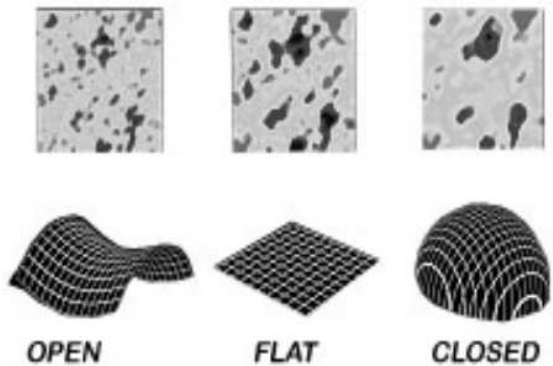}}   
\caption{The geometry of the universe can be read off from the angular spread of the dominant temperature anisotropies in the cosmic microwave background. In this picture dark patches are regions with slightly higher temperature at the moment of last scattering 300 ky after the Big Bang. Their angular spread would appear to us larger or smaller in a non-euclidean universe. \label{fig3}}
\end{figure}

Until the last scatterings of photons with matter occurred, photons, electrons and baryons formed essentially a baryon-photon plasma. Fluctuations in this "fluid" with locally increased baryon density had a tendency to be amplified by the pull of gravity, but at the same time radiation pressure built up trying to decrease the baryon density. This led to acoustic oscillations of the "baryon-photon fluid". If a compression node happened to occur at the moment of last scattering, then the heated plasma led to hot spots also in the CMB. What maximum size could these regions occupy on the last scattering surface? Certainly they could not grow larger than the speed of sound times the maximum time of propagation since the Big Bang, i.e. 300 ky!  Today, 13 billion years later, we should see these regions on the sky as patches, subtending an angle of 0.9$^\circ$ if we live in a flat universe, where the lines of sight are straight. Smaller or larger patches would be evidence of a non-zero curvature of the universe (fig. \ref{fig3}). WMAP tells us that the dominant patch size is indeed 0.9 degree. Conclusion: we live in a flat universe with\cite{HIN} $\Omega_{tot}$ = 1.003 +0.013/-0.017 ! \\ 

A detailed statistical analysis of the patch sizes in the temperature anisotropies shows a series of peaks on decreasing angular scales, which are related to various harmonics in the oscillation modes. Their location and relative amplitudes gives us a wealth of information, with the following results: $\Omega_\Lambda$ = 0.73 $\pm$ 0.04, $\Omega_m$ = 0.27 $\pm$ 0.04, $\Omega_b$ = 0.044 $\pm$ 0.004, a Hubble constant of $H_0$ = 71 $\pm$ 4 km s$^{-1}$Mpc$^{-1}$ and an age of the universe of $T_0$ = 13.7$\pm$0.02 Gyr.\\

Where did these tiny, but most significant fluctuations in the baryonic density come from? Probably from quantum fluctuations in the very early universe. Of course, inflation largely reduced their amplitudes and spread them out spatially, but they survived. They seeded the acoustic oscillations of the early baryon-photon plasma and gave rise to the observed CMB anisotropy.\\   

\section{Dark Matter and the Development of Structure in the Universe}

\begin{figure}[ht]
\centerline{\epsfxsize=4.1in\epsfbox{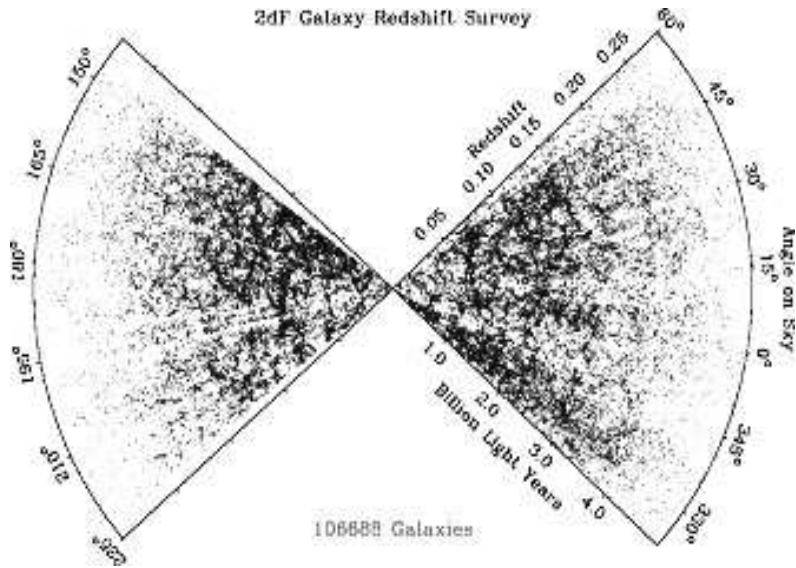}}   
\caption{Deep space galaxy surveys like this one executed by of the 2dF collaboration have shown that galaxies are not randomly distributed, but rather form Large Scale Structures, with patterns of foam like filaments surrounding vast regions of empty space. Calculations show, that such structures can be explained by a much more important distribution of cold dark matter particles, which we can trace now with luminous galaxies. \label{fig4}}
\end{figure}

If we look at an image of the distribution of far away galaxies, like the one shown in fig. \ref{fig4}, we notice that galaxies are not scattered in a random way, but form a foamy network of filaments, strings of clusters and sheets, sometimes surrounding huge regions of empty space. More than half a million galaxies have been mapped by the Sloan Digital Sky Survey, the 2dF Galaxy Red shift Survey and others, in a vast region of space, which covers a cube of 6 BLy a side. This so-called large-scale structure cannot be explained as a result of gravitational clumping of baryonic matter by itself. \\

We know that baryonic nucleosynthesis occurred about two hundred seconds after the Big Bang, which is too late for any appearing structures to be not washed out by the Hubble expansion. But assume the much more abundant, weakly interacting dark matter decoupled much earlier from thermal equilibrium, then it had more time to develop structure and therefore could clump earlier. Then, after the time of last scattering, when CMB photons ceased to interact with baryonic matter, the latter could fall into the gravitational troughs at locations with high dark matter concentrations and form galaxies. This explanation solves three problems at once: it explains galaxy formation in the right mass range, the observed large-scale structure and gives us another clue about the nature of dark matter.\\

In order to explain the observed features of large scale structure with dark matter some kind of non baryonic matter, in the form of neutral particles, weakly interacting particles (WIMPs) could fit the bill. Neutrinos would be also candidates, but with the tiny masses they have, they would be relativistic. This kind of "hot" dark matter can be shown in simulations to favour clustering at larger scales than observed. On the contrary, non-relativistic, slow heavy neutral particles, "cold" dark matter, would cluster on small scales and would develop later structures on larger scales. This so-called bottom-up model fits well the observed overall matter distribution and predicts galaxy formation in the right mass range from 10$^{-3}$  - 10$^4$ of the mass of our Milky Way.\\

\begin{figure}[ht]
\centerline{\epsfxsize=3.0in\epsfbox{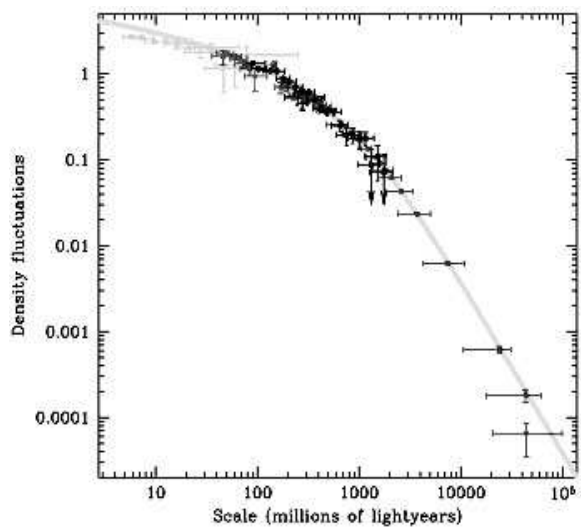}}   
\caption{By analysing the large scale structure of galaxy distributions in terms of density fluctuations, one finds that the universe becomes more and more uniform at larger scales and the power spectrum is best described by a $\Lambda$-CDM model compatible with the cosmological parameters of WMAP\protect\cite{TEG}. \label{fig5}}
\end{figure}

In fact, the analysis of galaxy density fluctuations has become another important tool to determine the cosmological parameters. The way to proceed is to take the Fourier transform of the two-point density correlation function $\delta (x)$ and to analyse the resulting power spectrum
\begin{equation} 
\delta (x) ~=~ \frac{\rho(x)-\bar{\rho}}{\bar{\rho}} ~=~ \sum_k \delta_k(k)e^{ikx}~~~~~~~~~~P(k)=|\delta_k|^2
\label{eq:eq12}
\end{equation}

where $\bar{\rho}$ is the mean galaxy density and $\rho (x)$ the density at a distance $x$ with respect to the barycentre of the averaged region The best fit to the power spectrum $P(k)$ is obtained for $\Omega_b$ = 5\%, $\Omega_m$ = 25\% and $\Omega_\Lambda$ = 70\%, in excellent agreement with the WMAP results. We also see from fig. \ref{fig5} that the larger the scale we average over, the more uniform the universe becomes, in agreement with the cosmological principle\cite{TEG}. \\

\begin{figure}[ht]
\centerline{\epsfxsize=4.1in\epsfbox{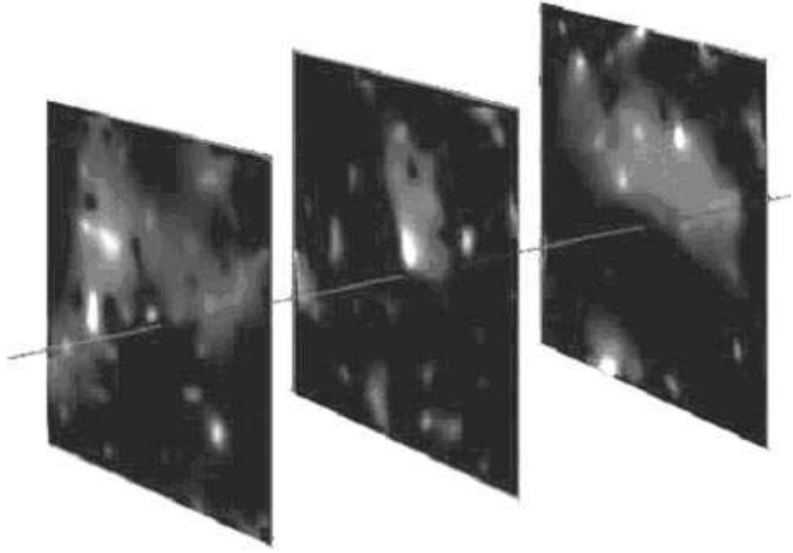}}   
\caption{The first 3D distribution of Dark Matter was reconstructed by gravitational lensing by the COSMOS collaboration. Shown are three redshift slices showing from left to right dark matter clumping 3.5 By ago, 5.0BY and 6.5By ago. As can be seen clumping increases with increasing age of the universe.  \label{fig6}}
\end{figure}

Do galaxies really trace the distribution of dark matter? The Cosmic Evolution Survey of the COSMOS collaboration investigated this question by reconstructing the first large-scale 3D image of a distribution of dark matter. For this to happen, the Hubble space telescope took the largest picture mosaic ever of the sky in the near infrared (1.4$^\circ$ x 1.4$^\circ$) and gravitational weak lensing was used to trace the location of dark matter over a region of 100x100 MLy$^2$. The distances to individual galaxies were provided by red shift measurements of the ESO / Magellan telescopes in Chile and the Subaru / CFHT telescopes on Hawaii; the XMM Newton observatory provided an X-ray map of the hot gas around galaxies. The results published in January 2007 show that dark matter appears to be more than six times more abundant than luminous matter and indeed the distribution of luminous matter and of hot gas follows closely that of dark matter\cite{RM}. Moreover  the information on redshift could be used to reconstruct the clumping of ordinary matter and dark matter at three different time slices: 3.5, 5.0 and 6.5 billion years ago. Most interestingly, fig.\ref{fig6} shows, that with increasing time during the Hubble expansion also the dark matter "lumpiness" grows, with ordinary matter flowing into the gravitational troughs. 

The future Large Synoptic Telescope starting in 2013 with its 8.4 diameter mirror will continue these studies with unprecedented precision, collecting 30 Tbyte of data per night. Google will participate in organizing the data crunching\cite{NS}.\\

\section{The Two Dark Matter Problems}

From our discussion above we learned that the universe is composed of 73\%  dark energy, 23\% of cold dark matter and 4\% of baryonic matter. Luminous matter contributes a mere 0.4\% to the total. Comparing these numbers, we must conclude that there are in reality two dark matter problems to be solved. Number one: most of the ordinary matter is dark! i.e. part of the missing mass in galaxies and clusters should be explainable by baryonic matter. Number two: most of the matter is non- baryonic! In fact, 85\% of gravitationally traceable matter (by weak lensing etc) should be non-baryonic!\\

\underline{Problem number 1}: If most of the ordinary matter in galaxies as our own is dark, what can it be? It could be brown dwarfs for example. They are known to exist, but there is no evidence that they can be nearly as abundant to explain the observed deficit in baryonic matter. Extensive searches have been carried out by the MACHO, EROS, OGLE collaborations to find so called MACHOS (Massive, Compact Halo Objects), by an occasional amplification of light by gravitational lensing of background stars. Several million stars were observed for years. Four candidates have been found towards the Large Magellan cloud, 45 towards the galactic centre. However these findings suggest that less than 20\% of the galactic dark halo can be explained by MACHOS in the range of 2x10$^{-7}$ to 1 solar mass.\\

It could be white dwarfs, but their observed abundance in the halo is less than 5\% according to observations with the Hubble DFD telescope. Also a sufficiently large number of white dwarfs requires more He than was produced during Big Bang nucleosynthesis. It could be still hydrogen gas, which is known to contribute 75\% of the visible mass in galaxies, but it is difficult to hide more. Neutron stars or black holes would also make good candidates, but they are even scarcer than white dwarfs and the processes, which produce them, release large amounts of energy and heavy elements for which there is no evidence.\\

A completely different approach to solve the dark matter problem at galactic and also larger scales was proposed by Mordechai Milgrom in 1983: Modified Newtonian Dynamics or MOND\cite{MM}! This empirical theory forces in a pragmatic way the rotation curves of galaxies to become flat. For this, Newton's 2$^{nd}$ law had to be modified for very small accelerations falling below a certain $a_0$, such that the actual accelerations become {\em larger} than Newtonian:  
\begin{equation}
\vec{F}~=~ \left\lbrace \begin{array}{lc} m\vec{a} & ~~~~~ {\rm for}~|\vec{a}|\gg |\vec{a}_0| \\ m\vec{a}\frac{|\vec{a}|}{|\vec{a}_0|} & ~~~~~{\rm for}~|\vec{a}|\ll |\vec{a}_0|\end{array} \right.
\label{eq:eq13}
\end{equation}

Therefore disc-stars at large distances from the bulge in the $a_0$ regime would obey a modified force-law
\begin{equation}
F ~=~ m\frac{a^2}{a_0} ~=~ G\frac{mM}{r^2}
\label{eq:eq14}
\end{equation}

Since on a circular orbit $a$ = $v^2$/$r$ the rotation speed $v$ becomes constant and independent of the distance from the galactic centre:
\begin{equation}
v ~=~ \sqrt[4]{GMa_0}
\label{eq:eq15}
\end{equation}

This relation resembles strikingly the famous Tully-Fisher relation between the rotation velocity of disc stars and the luminosity of a galaxy. But on the other hand the theory was constructed precisely to give this result! Knowing $v$ and $M$ one can calculate $a_0$ and obtains $a_0$ = 1.2x 10$^{-10}$ ms$^{-2}$, which curiously is close to $cH_0$ or the speed of light divided by the age of the universe! \\

The success of MOND to get rid of dark matter in certain situations has sparked the formulation of more generalized, relativistic versions of MOND, like the Tensor-Vector-Scalar Theory (TeVeS) of J.D. Beckenstein\cite{BECK}, conformal gravity by P.D. Mannheim\cite{MANN}, and non-symmetric gravitational theory by J. W. Moffat\cite{MOFF}.  But there are also difficulties with MOND and its derivatives. The non-relativistic theory violates momentum conservation, Lorentz invariance and the equivalence principle. The theory is not self-consistent when it comes to extending ideas beyond individual galaxies\cite{DSC}. The observed behaviour of matter and dark matter in the bullet cluster, can only partially be explained by MOND (in fact 2eV neutrinos are needed in addition); to explain weak gravitational lensing of clusters of galaxies and large scale structure relativistic versions like TeVeS are needed, which introduce important additional complexities and problems; the CMB anisotropy is explained with great difficulty; $a_0$ is not a unique constant but can vary from galaxy to galaxy up to factors of five; fits to clusters of galaxies require higher values for $a_0$ then for individual galaxies; dwarf galaxies need dark matter despite of MOND; finally there are galaxies which show Keplerian velocity profiles, which are completely incompatible with any MOND theory. Nevertheless, in view of the successful predictions at the galactic scale it is certainly very interesting to explore if MOND is capable to deliver reliable predictions at larger, cosmic scales.\\

\underline{Problem number 2}: Since baryonic matter is about a factor 6 less abundant than all the gravitating matter together, most of matter must be of some yet unidentified non-baryonic kind. The most popular hypothesis is that non-baryonic dark matter consists of some neutral massive weakly interacting particles (WIMPS), which were created in the hot early universe, decoupled early from ordinary matter in order to seed structure formation, as discussed in chap. 4 and survived until today. There is a plethora of candidates: it could be neutrinos, axions, Kaluza-Klein gravitons, gravitinos, neutralinos, sneutrinos, primordial black holes, particles from little Higgs models etc. How do we find the right kind? \\

First of all we must make sure that these particles can have survived until today with the right abundance, consistent with $\Omega_m$ = 0.23. So let us pick a WIMP candidate $x$ together with its antiparticle $\bar{x}$ and lets suppose that they can annihilate each other and be created in pairs via reactions like $x~+~\bar{x}~\rightarrow~f~+~\bar{f}$ where $f$ and $\bar{f}$ are particles and their antiparticles e.g. pairs of leptons or quarks. Let us assume further that in the hot early universe $f$ and $\bar{f}$ were in thermal equilibrium with photons and all the other light particles. How would the number density $n_x$ of our WIMP evolve in time? The answer is given by the Boltzmann equation:
\begin{equation}
\frac{dn_x}{dt} ~=~ -3Hn_x ~ - ~n_x^2<\sigma_{x\bar{x}\rightarrow f\bar{f}}~v>~+~n_f^2<\sigma_{f\bar{f}\rightarrow x\bar{x}}v~>
\label{eq:eq16}
\end{equation}

The first term on the right side describes the dilution of the WIMPs by the Hubble expansion, the second the depletion rate by annihilation and the third term the rate of creation by pair production. Both terms are proportional $n^2$ since particle and antiparticle densities are equal and are multiplied by the thermally averaged relative speeds and cross-sections. In thermal equilibrium the last two terms are equal and the number density of WIMPS is the equilibrium density $n_x^{eq}$. Any departure from equilibrium is described by 
\begin{equation}
\frac{dn_x}{dt} ~=~ -3Hn_x ~ - ~(n_x^2~-~n_x^{eq,2})<\sigma_{ann}v>
\label{eq:eq17}
\end{equation}

where $\sigma_{ann}$ is the total annihilation cross section\cite{KT,BG}. The equilibrium density depends on the temperature in the early universe and the mass of the considered particle type. At high temperatures exceeding the WIMP mass $m_x$ we have $n_x$ = $n_{eq}$. When the temperature drops below the WIMP mass $m_x$, the pair creation would require ordinary matter particles from the tail of the thermal velocity distribution. Therefore in equilibrium the number density falls off exponentially with
\begin{equation}
n_{eq} ~\propto~ (mT)^\frac{3}{2}e^{\frac{-m_x c^2}{kT}}
\label{eq:eq18}
\end{equation}

If the particles would continue to remain in thermal equilibrium, few would be left over today. But at the moment when the annihilation rate $n_x <\sigma_{ann}v>$ becomes smaller than the expansion rate $H$, the probability of WIMP particles to find a partner for annihilation will become small. The WIMP number density "freezes out" and can survive until today. After integrating eq.(\ref{eq:eq17}) we can calculate the number density $n_x(t_0)$ of today and the expected mass parameter for our dark matter candidate becomes:
\begin{equation}
\Omega_x ~=~ \frac{n_x(t_0)m_x}{\rho_c(t_0)} ~=~ \frac{3{\rm x}10^{-27}{\rm cm}^2{\rm sec}^{-1}}{<\sigma_{ann}v>h^2}
\label{eq:eq19}
\end{equation}

\begin{figure}[ht]
\centerline{\epsfxsize=3.0in\epsfbox{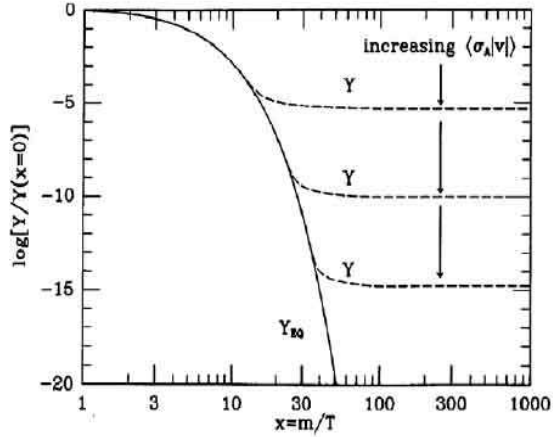}}   
\caption{The WIMP abundance in thermal equilibrium is decreasing with decreasing temperature. At a certain moment, when the Hubble expansion rate exceeds the annihilation rate, annihilation partners get too far separated and  WIMPs are "freezing" out. The larger the annihilation cross section is, the longer the particles remain in equilibrium and the smaller will be the relic abundance.    \label{fig7}}
\end{figure}

The dependence on today's Hubble constant $H_0$ enters in the form of the so-called "Hubble parameter" $h$ = 0.73 +0.03 / -0.04 which is defined via the relation $H_0$ = $h$ 100 km s$^{-1}$Mpc$^{-1}$. The WIMP abundance as a function of the dimensional parameter $x$ = $m$/$T$ i.e. with increasing time, is shown in fig. \ref{fig7} from ref. \refcite{KT}. Now all what remains to be done is to get all the channels the candidate particle can annihilate into, calculate $\Omega_x$ and compare it with the current preferred value of $\Omega_m$ $\sim$ 0.23. \\

Finally we draw several interesting conclusions: i) a valid WIMP particle must be a stable particle in order to survive; ii) For $\Omega_x$ to fall into the right range we need a particle which interacts with about electro-weak strength; iii) the larger the total annihilation cross section is, the longer the WIMPs remains in thermal equilibrium and the smaller their relic abundance will be!\\

\section{Non-baryonic Dark Matter Candidates}

Neutrinos would be, in principle, excellent candidates for non-baryonic dark matter. They exist, they are neutral and interact weakly. Their relic abundance\cite{COW} can be related to their mass by
\begin{equation}
\Omega_\nu ~=~ \frac{\sum_i m_{i\nu}}{50{\rm eV}}
\label{eq:eq20}
\end{equation}

Therefore, in order to fall into the interesting range of, say 0.17 $<$ $\Omega_\nu$ $<$ 0.25, the sum of all neutrino masses should be in the gamut of 8eV $<$ $\sum m_\nu$ $<$ 12eV. But the Mainz-Troitsk spectrometer experiment gives us an upper limit on the mass of the electron neutrino of $m_{\nu_e}$ $<$ 2.05 eV (90\% c.l.) only.  The experimental limits on the muon and tau- neutrino masses are much weaker. But if we take into account the observed mass squared differences from neutrino oscillation experiments with $\Delta m^2$ = 7x10$^{-5}$ eV$^2$ and $\Delta m^2$ = 3 x10$^{-3}$ eV$^2$ for solar and atmospheric neutrinos, respectively, we expect that 2 eV will be pretty much the ceiling for the sum of neutrino masses. If we add, that CMB and Large Scale Structure leave room only for neutrino masses with $m_\nu$  $<$ 0.23 eV, then we must conclude, unfortunately, that neutrinos fall short of hitting the mark. \\

Axions are very special cold dark matter candidates. They have been invented for theoretical reasons in order to explain the absence of CP violation in strong interaction processes and might have been created in the early universe, where they obtained their mass during the QCD phase transition when the quark-gluon plasma condensed into hadrons. In contrast to the WIMP freeze out mechanism described above, axions were never in thermal equilibrium. At decoupling they formed a bose condensate of particles with zero momentum. The axion mass $m_a$ is related to the energy scale of the phase transition $f_a$ by 
\begin{equation}
m_a ~=~ 0.62{\rm eV}\frac{10^7 {\rm GeV}}{f_a}
\label{eq:eq21}
\end{equation}

Axions are pseudo-scalar particles, which interact with nucleons, electrons and photons with a coupling constant, which goes like $g_a$ $\propto$ $f_a^{-1}$. If axions couple directly to the electron, one speaks of a DFSZ axion, if it couples indirectly over higher order corrections to the electron it is a KSZV axion. For $f_a$ $<$ 10$^{12}$ GeV, axions behave like cold dark matter particles.  They interact extremely weakly with matter and with masses in the range of 10$^{-6}$ eV $<$ $m_a$ $<$ 10$^{-4}$ eV they are able to generate a density parameter in the range of 0.1 $<$ $\Omega_m$ $<$ 1. \\

\begin{figure}[ht]
\centerline{\epsfxsize=4.1in\epsfbox{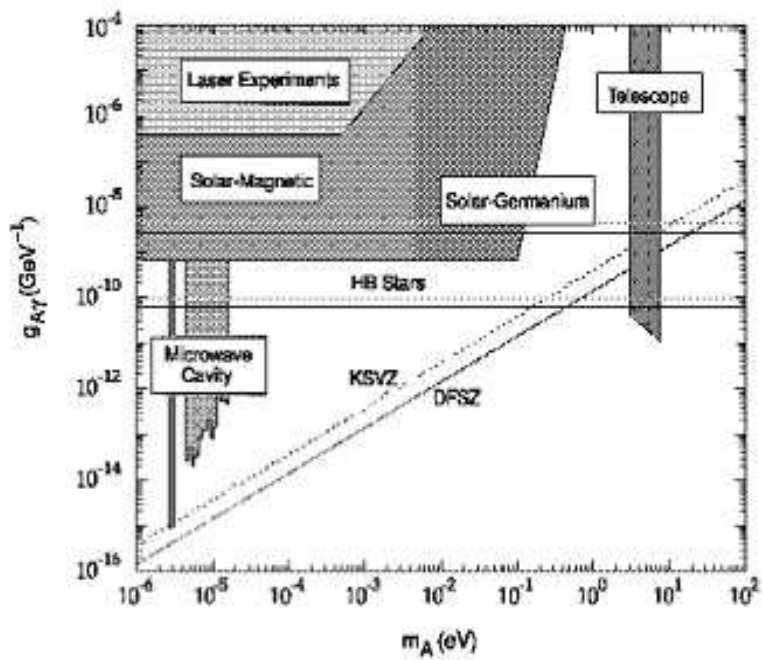}}   
\caption{Summary of experimental limits on the axion detection. $g_{A\gamma}$ is the coupling constant of axions to photons. Shown are the limits of the ADMX microwave cavity experiment. The solar axion experiment CAST is quoted as "solar magnetic". The two diagonal lines denoted KSVZ, DFSZ represent axions, which are valid candidates for cold dark matter in the galactic halo\protect\cite{AZ1}. \label{fig8}}
\end{figure}

The fact that axions couple to electromagnetic fields is used in experiments to detect them. The Axion Dark Matter Experiment (ADMX) is the first experiment with sufficient sensitivity to probe dark matter axions in the galactic halo. The axions are detected by the Primakoff effect, i.e. they are made to interact with the virtual photons of an 8 Tesla strong B-field and are converted into photons within a cryogenic, tuneable microwave cavity. A signal would show up as an excess power in the cavity if the mode frequency were close to the mass of the axion.  A mass range of 2 -3 $\mu$eV was scanned and with more than 90\% confidence the experiment could rule out KSVZ axions as halo particles, but could only weakly exclude a dark matter halo composed of DFSZ axions\cite{AZ}. Fig. \ref{fig8} gives a summary of experimental limits\cite{AZ1} as a function of mass and coupling constant.\\

The CERN Axion Solar Telescope (CAST) searches for axions created by the Primakoff effect in the sun's interior and detects them in the 9 Tesla strong magnetic field of the LHC magnets, following the sun's daily movement, again by converting axions to x-ray photons. The experiment is sensitive to axion masses up to $m_a$ =0.02 eV. After 3 years of data taking no axions have been yet detected. The limits are labelled as "solar magnetic" in fig. \ref{fig8}. \\

In 2006 the PVLAS experiment at INFN Legnaro reported an intriguing result, which could be interpreted as evidence of an axion-like particle. In their apparatus laser light passing through a strong magnetic field showed a small rotation of its polarisation plane. The observed effect was four orders of magnitude larger than predicted by QED and could be explained by the production of an axion-like particle which attenuates the E-field component of the laser light parallel to the B-field.  The mass of this particle and its coupling strength to photons would be around\cite{AH} 1 meV and 2x10$^{-6}$ GeV$^{-1}$. Such a particle could however not be a QCD axion, because according to eq. \ref{eq:eq21}, it should have couplings, which are at least seven orders of magnitude smaller. Besides PVLAS a series of new experiments like ALPS, BMW, LIPPS are starting taking data in 2007 to verify the existence of the claimed new particle. On the theoretical side, there may be some possibilities to explain such an axion-like particle in the framework of milli-charged fermions or axionic superstrings, but a major difficulty resides in the fact that these axion- like particles should be scalars, whereas axions are pseudo-scalars.\\

\section{Neutralino Cold Dark Matter}

From the particle physicist's point of view, neutralinos are, together with axions, the best-motivated candidates for non-baryonic cold dark matter\cite{BER}. Neutralinos, often referred to as $\chi$, can form the lightest stable supersymmetric particle (LSP) into which all heavier SUSY particles decay if R-parity is conserved. Their interaction with matter is electro-weak. Neutralinos are heavy, with a mass range of 45 GeV $<$ $M_\chi$ $<$ 7 TeV, with the lower limit set by LEP and the upper limit given by cosmology. They can form a relic population early after the Big Bang and can provide an $\Omega_m$ in the right range. Relic $\chi$'s  are non-relativistic and can explain the development of large-scale structure. \\

The neutralino is supposed to be the lightest linear combination of the supersymmetric partners of the neutral gauge bosons and neutral Higgses, i.e. the Photino, Zino and Higgsinos:
\begin{equation}
\chi_1 ~=~ N_{11}\tilde{\gamma} ~+~N_{12}\tilde{Z}  ~+~N_{13}\tilde{H}_1^0 ~+~N_{12}\tilde{H}_2^0
\label{eq:eq22}
\end{equation}

The character of the particle, i.e. if it is more Higgsino- or more Gaugino- like and the kind and strength of interaction with ordinary matter depend on the parameters of  the underlying supersymmetric model. Unfortunately, already the Minimal Supersymmetric Model (MSSM) has more than 100 parameters. Therefore the following strategy is usually taken:

\begin{romanlist}
\item Select a model with a reduced set of parameters, like minimal Supergravity (mSUGRA), pMSSM...

\item Apply experimental constraints on the SUSY parameter space derived e.g. from lower bounds on the Higgs and Chargino masses, from results of $b\rightarrow s\gamma$ and the measurements of $g_\mu - 2$.

\item Get the total cross section for annihilation into all possible channels and calculate the relic density according to eq. \ref{eq:eq19}.

\item Check if $\Omega_m$ falls into the expected range predicted by WMAP.

\item If yes, take a code like DARKSUSY\footnote{can be downloaded from http://www.physto.se/$\sim$edsjo/ds} in order to calculate the neutralino - proton cross sections for dark matter experiments

\end{romanlist}

If we take for example mSUGRA, also called minimal or constrained MSSM, we end up with four parameters and one sign\cite{BAL,ELL}. The parameters are tan$\beta$ = $<H_2>$/$<H_1>$, the ratio of Higgs vacuum expectation values, M$_{1/2}$ is the unified Gaugino mass at the grand unification scale, M$_{GUT}$; $m_0$ is the  unified scalar mass at $M_{GUT}$ (with roughly $M_\chi$ $\propto$ $m_\frac{1}{2}$).  $A_0$, describes the trilinear coupling strength in the SUSY Lagrangian and is usually set to zero. Finally sign ($\mu$), is the sign of the Higgsinos mass parameter $\mu$.\\

If we calculate now $\sigma_{ann}$ for all possible channels by varying our model parameters, we find that depending on the neutralino type, $\sigma_{ann}$ can become small if it's a Bino type neutralino, large for a Higgsino and huge for a Wino type particle. Correspondingly the density parameter $\Omega_m$ becomes large for a Bino, small for a Higgsino and tiny for a Wino type neutralino. In fact for a Bino type Neutralino, $\Omega_m$ would come out much too large for most of the parameter space. However it happens that if e.g. some sfermions are close or degenerate in mass with the neutralino mass, "co-annihilation" can occur, which increases $\sigma_{ann}$  in a resonance like manner. Especially the co-annihilation with the superymmetric tau lepton, $\chi + \tilde{\tau}\rightarrow\tau +\gamma$ would be a probable process. \\

\begin{figure}[ht]
\centerline{\epsfxsize=3.0in\epsfbox{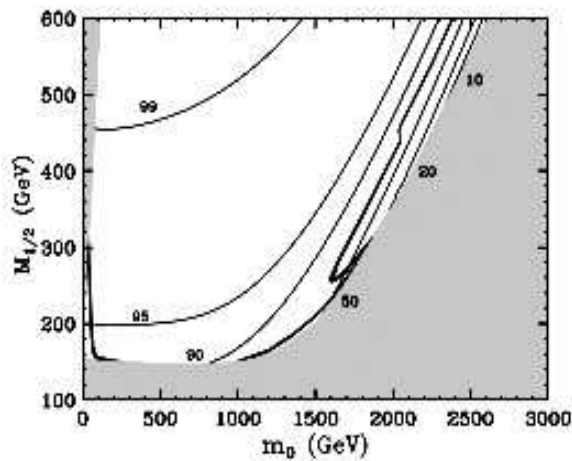}}   
\caption{Parameter space $M_{1/2}$ vs. $m_0$ for mSUGRA parameters tan$\beta$ = 10, $A_0$ = 0 and  $\mu$ $>$ 0. The grey region is excluded by accelerator and theoritical constraints. After applying the cosmological constraints of WMAP only two too tiny regions remain allowed: the dark lined region to the left at small $m_0$ (co-annihilation region) and the dark lined region at high $m_0$ following the grey contour line (focus point region)\protect\cite{KTM}. \label{fig9}}
\end{figure}

The allowed parameter space $M_{1/2}$ vs. $m_0$, which survives after application of the cosmological constraints is shown in fig. \ref{fig9} for mSUGRA parameters\cite{KTM} tan$\beta$ = 10, $A_0$ = 0 and $\mu$ $>$ 0. Two tiny regions remain where the neutralino is a valid dark matter candidate: the so-called "focus-point" region at large m$_0$ where the neutralino has a large Higgsino component and the "co-annihilation" region for small $m_0$, where the neutralino is gaugino-like. Similar allowed regions, although shifted are generated for other choices of parameters.\\

Having identified the remaining, allowed SUSY parameter space, we can calculate next the cross sections for neutralino interactions with matter\cite{TOV,BEL}. The interactions can be axial, i.e. spin dependent (via Z or squark exchange) or can be  scalar, i.e. spin-independent (via H or squark exchange) or both. This depends again on the neutralino type! The general form of the cross section for a neutralino interaction with a nucleus of atomic number A has the form:
\begin{equation}
\sigma_A ~=~ 4G_F^2 \left( \frac{M_\chi M_A}{M_\chi +M_A} \right)^2 C_AF(q^2)
\label{eq:eq23}
\end{equation}

$C_A$ is an enhancement factor, which depends on the type of the interaction. $F(q^2)$ is a nuclear form factor, which becomes important only for large $A$ and large momentum transfers. For spin-independent interactions $C_A$ is described by 
\begin{equation}
C_A^{SI} ~=~ \frac{1}{4}\left( Zf_p ~+~ (A-Z)f_n \right)^2 
\label{eq:eq24}
\end{equation}

where $f_{p,n}$ describes the coupling to the nucleons. If  $f_p$ = $f_n$,  the interaction is enhanced by $A^2$. However both couplings can also interfere destructively, leading to very small cross sections. In order to compare the theory with experiment, and also different experiments among each other using different nuclei, one normalizes usually to the neutralino-proton cross section  
\begin{equation}
\sigma_p^{SI} ~=~ \frac{1}{A^2}\left( \frac{\mu_p}{\mu_A} \right) \sigma_A
\label{eq:eq25}
\end{equation}

Here $\mu_{p,A}$ is the reduced mass of neutralino-proton and neutralino-nucleus. Similarly we can write down a spin-dependent enhancement factor
\begin{equation}
C_A^{SD} ~=~ \frac{8}{\pi} \left[ a_p<S_p>~+~a_n<S_n> \right]^2 (J+1)J
\label{eq:eq26}
\end{equation}

where $J$ is the spin of the nucleus and $a_{p,n}$ are the coupling constants on protons and neutrons. $<S_{p,n}>$ are the averaged spins over all protons and neutrons in the nucleus, respectively. Writing eq. \ref{eq:eq26} more compactly, we get $C_A^{SD}~\propto ~\lambda^2 J(J+1)$. Only a few nuclei have a large spin-dependent cross section. Next to the bare proton ($\lambda^2$ = 1), the most favourable nucleus for spin-dependent interactions on protons is $~^{19}F$ ($\lambda^2$ = 0.86). Popular nuclei in dark matter search detectors like $~^{23}$Na and $~^{127}$I have much smaller sensitivities ($\lambda^2$ = 0.011 and 0.0026, respectively).  As in the spin-independent case, nuclear cross sections are normalized to the respective nucleon cross sections:
\begin{equation}
\sigma_{p,n}^{SD} ~=~ \frac{3}{4}\frac{J}{J+1}\left( \frac{\mu_{p,n}}{\mu_A} \right)^2 \frac{1}{<S_{p.n}>}\sigma_A^{SD}
\label{eq:eq27}
\end{equation}

\begin{figure}[ht]
\centerline{\epsfxsize=4.1in\epsfbox{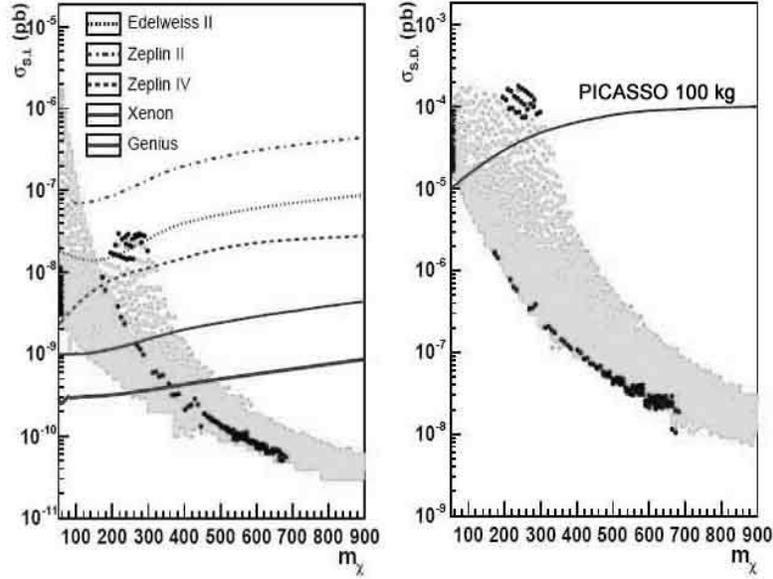}}
\caption{ Predicted cross-sections in the spin independent (left) and spin dependent sector for a mSUGRA model with non-universal gaugino masses. The model parameters are m$_0$ $<$ 5 TeV, M$1/2$ $<$ 2TeV, $\mu$ $>$ 0 , A$_0$ = 0 and with a top mass of m$_t$ = 175 GeV. Broken lines are present experimental limits. Full lines are projected limits.  \label{fig10}}
\end{figure}

How do spin-dependent and spin-independent cross-sections compare? Unfortunately we do not know yet what choice (if ever) nature made. So the best we can do is to get an estimate of what values of cross sections in each channel are compatible with the allowed parameter space for the model we have chosen. One of the results is shown in the scatter plots of cross section on protons versus neutralino mass in fig. \ref{fig10}. The theoretical framework is a model were Gaugino masses were allowed to vary for m$_0$ up to 2 TeV and m$_\frac{1}{2}$ up to 10 TeV, respectively and for different values of tan$\beta$ (only tan$\beta$~=~50 is shown here)\cite{BEL}. As can be seen, in both cases allowed neutralino masses range from about 100 GeV up to 900 GeV. Spin-independent cross sections range from 10$^{-6}$ to 10$^{-11}$ pb and in the spin-dependent case from 10$^{-4}$ to 10$^{-8}$ pb. Why then, as we will see later, do most experiments search for spin-independent interactions? Well, a heavy target gives a larger sensitivity because of the $A^2$ dependence of the cross section, which more than compensates the smaller a priori cross section on individual nucleons. \\

\begin{figure}[ht]
\centerline{\epsfxsize=4.1in\epsfbox{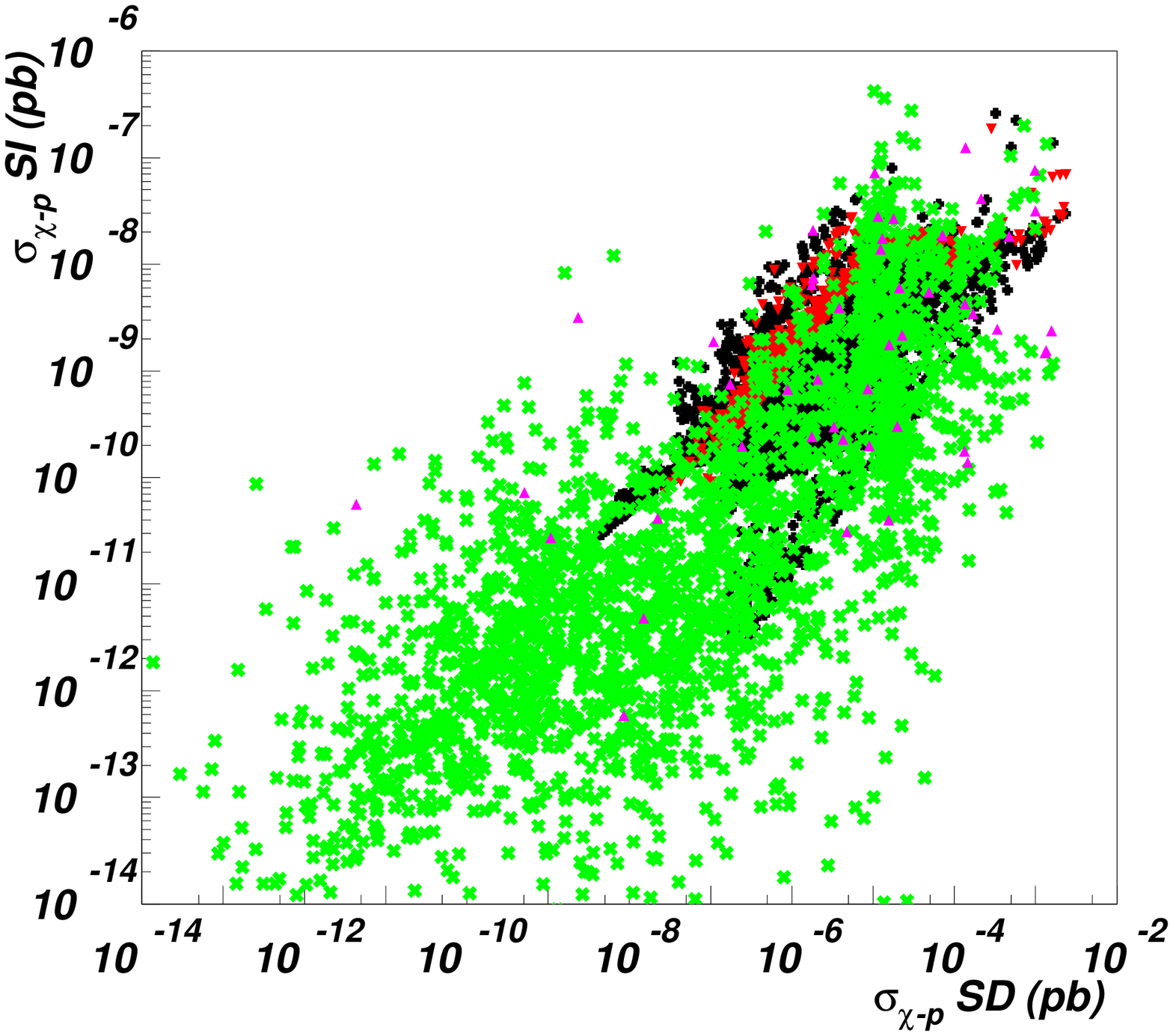}}   
\caption{Results of a correlation\protect\cite{MHG} study of spin-dependent versus spin-independent cross-sections in a model of MSSM without unification conditions where Gaugino and sfermion masses vary up to 10 TeV, pseudoscalar Higgs mass up to 1 TeV and tan$\beta$ from 1 to 60. When the predicted relic density was too low, i.e. $\Omega_\chi h^2\leq$0.095, the cross-section was scaled by a factor  $\Omega_\chi h^2$/0.095 in order to allow direct comparison with the experimental limits on the cross sections. Large spin-dependent values can correspond to small spin-independent cross sections. \label{fig11}}
\end{figure}

However, it could still be that neutralinos favour spin-dependent interactions. Therefore in order to figure out where we should look, we have to investigate whether there are correlations between the spin-dependent and spin-independent cross sections. This has been done in ref. \refcite{BOT} and \refcite{MHG} and the result shows that both channels are quite loosely correlated. It can happen e.g, that a spin-dependent cross section of 5x10$^{-3}$ pb corresponds to a spin-independent cross section of 10$^{-11}$ pb.  Therefore a dark matter signal can be missed eventually, if searches are carried out in one channel only!\\

\section{Searches for Non-baryonic Cold Dark Matter}

There are three ways to search for particle Dark Matter. Direct searches try to detect interactions with galctic WIMPs in the laboratory. As we will see below, with an average speed in the halo of 270 km/sec these are fast particles, which can produce measurable recoils in a detector. These experiments probe the neutralino halo density and the halo composition at the location of the solar system. If no signal is observed, experiments can give at least limits on the interaction cross-section and the MSSM parameter space.\\

	In indirect searches it is assumed that WIMPS can be trapped gravitationally in the galactic centre or the interior of stars like our sun. Being Majorana particles neutralinos can annihilate with each other e.g. into pairs heavy quarks, W and Z-bosons and the following energetic gamma rays, neutrinos and muons are detected on earth. These experiments are able to probe the halo composition also elsewhere, like in the galactic centre or in the entire halo. Again the absence of a signal is still a useful information in order to set limits on cross sections and theoretical parameters.\\

The search for SUSY particles is one of the main motivations of the upcoming LHC at CERN and the search for WIMP candidates is at the centre of everyone's interest. If no signal is found, at least the limits on neutralino masses can be pushed from 50 GeV to 300 GeV. A future linear collider like the ILC might be able to explore mass ranges up to 400 GeV\cite{SCH,ARN}.\\

Is one search more promising than the other? In fact direct, indirect and accelerator searches are complementary! The discovery of a cosmological WIMP does not yet prove Supersymmetry; this has to be confirmed by accelerator experiments. On the other hand an LHC signal cannot prove that the candidate particle detected by missing energy is a stable Dark Matter candidate. For this, direct and indirect experiments are needed. \\ 

\section{Status of Indirect Searches}

WIMPs might be gravitationally trapped in the sun's central region after loosing energy during many elastic scatterings with $\sigma_{SD}$ or $\sigma_{SI}$ on protons until they annihilate. After billions of years the annihilation and capture rates reach equilibrium with a capture rate of\cite{DHO}:
\begin{equation}
R({\rm sec}^{-1}) ~=~ 10^{18}\left( \frac{\sigma^p_{SD,SI}}{10^{-8}{\rm pb}} \right)\left( \frac{100{\rm GeV}}{M_\chi} \right)^2
\label{eq:eq28}
\end{equation}

But only neutrinos are able to escape the suns interior! Their flux will depend on the capture rate and especially on $\sigma^p_{SD}$ since the spin dependent cross section on protons is more important over a large region in the allowed SUSY parameter space. Event rate predictions for a km scale detector with a 50 GeV energy threshold can vary from\cite{HAL} 10$^2$ events/km$^{2}$/y to 10$^{-5}$ events/km$^{2}$/y. Among the experiments, which will search for solar annihilation neutrinos are AMANDA, ICeCube, ANTARES, NESTOR and Super Kamiokande.\\

Neutralino annihilation in the galactic halo into pairs of $b\bar{b}$, $W^+W^-$, $Z^0Z^0$ etc creates excess antimatter particles, especially positrons, which AMS-2 and PAMELA try to detect in space. The energy spectra of positrons reach from tens to hundreds of GeV. They depend on the annihilation mode and in a quite complicated way on positron diffusion in the galactic magnetic field, energy loss processes and the halo structure itself\cite{DHO}. Gamma rays from neutral pion decays can also be a signature for neutralino annhilation in the halo. The resulting gamma ray spectra range from a few GeV to several tens of GeV. Advantage: gammas are largely independent from the annihilation mode and can propagate freely without energy loss. Only the $\tau^+\tau^-$ channel differs in shape which allows a determination of the $\tau$ fraction from the spectral curve. \\

Gamma rays at these energies cannot be detected directly on earth. When those photons interact with the atmosphere, an electromagnetic cascade is created and the secondary shower particles can either be detected directly on ground or by their creation of Cerenkov photons during their passage through the atmosphere. Ground based Cerenkov telescopes for gamma ray searches are e.g. HESS, MAGIC, VERITAS and CANGOROO III. \\

Among these ground-based Cerenkov experiments, HESS has observed an excess of gamma rays coming from the galactic centre, which could be explained by 10 TeV mass Dark Matter particles, which challenges existing theoretical models. The CACTUS solar array has observed an excess of gamma rays at 100 GeV in the direction of the Draco dwarf galaxy, a companion of the Milky Way.  Also EGRET the first space based gamma ray observatory reaching in energy up to 30 GeV, observes an excess of diffuse gamma rays from the galactic halo at energies between one and 20 GeV.  Critics claim, that to explain this, a very peculiar composition of the galactic halo is required, with e.g. concentric rings of Dark Matter; also the theoretical models need important boost factors to explain the EGRET gamma flux. In any event, GLAST a next generation high energy gamma ray observatory which follows in the footsteps of the EGRET satellite and which is due to be launched in fall 2007, will hopefully shed light on these open issues. \\

Still another reported anomaly comes from the INTEGRAL satellite experiment, which explores the gamma ray sky in the keV to MeV region. It observed a strong signal of 511 keV photons originating from the galactic centre, which was interpreted as a tell tale signal of the annihilation of MeV scale WIMPs into electron-positron pairs.  However such particles would have been difficult to miss in accelerator experiments! Let's wait and see!\\

\section{Status of Direct Searches}

If today's Dark Matter paradigm is correct, our galaxy is surrounded by a spherical halo of self-gravitating WIMP particles, i.e. particles, which travel undisturbed in Keplerian orbits around the galactic centre with a Maxwellian velocity distribution.  The rotation curve of the Milky Way indicates that the halo particle density should fall off with distance from the galactic centre like 1/r$^2$ and in the vicinity of the solar system the mass-energy density should be around 0.3 GeV/cm$^3$. To the Maxwellian velocity distribution of the halo WIMPs with a dispersion of v $\approx$ 230 kms$^{-1}$ we still have to add the relative velocity of the solar system of 244 kms$^{-1}$ with respect to the halo. This give an average WIMP velocity around 240 kms$^{-1}$. How many halo-WIMP particles traverse each of us per second? Well, it comes to about 10$^{9}$!\\

The only way to detect WIMP interactions with matter is via their elastic scattering off a detector nucleus: following scattering with a WIMP of mass $M_\chi$ and energy $E_\chi$, a nucleus of mass $M_N$ will recoil at a scattering angle $\theta$ with an energy $E_r$   
\begin{equation}
E_r ~=~ 2E_\chi \frac{M_NM_\chi}{(M_N+M_\chi)^2}(1-cos\theta)
\label{eq:eq29}
\end{equation}

Therefore for monoenergetic WIMPS and isotropic scattering in the centre of mass system, we obtain the usual boxlike recoil spectra in the laboratory. If we then fold the Maxwellian kinetic energy distribution of the halo WIMPs with the above recoil distribution we get an exponentially falling recoil spectrum of the form
\begin{equation}
\frac{dR}{dE_r} ~\propto~ e^{-\frac{E_r}{<E_r>}}
\label{eq:eq30}
\end{equation}

here $<E_r>$ is the average recoil energy. For the parameters of our galactic halo we expect mean recoil energies of the order of a few keV to 100 keV following the relation
\begin{equation}
<E_r> ~=~ 2\left( \frac{M_A}{{\rm GeV}} \right) \left[ \frac{M_\chi}{M_\chi +M_A}\right]^2 ({\rm keV})
\label{eq:eq31}
\end{equation}

Therefore detectors with keV thresholds are required! Those low energetic recoil nuclei are notoriously difficult to detect and also difficult to discriminate against other kinds of low energy backgrounds. This is one of the many challenges WIMP detection experimenters have to face. \\

Another challenge is the interaction rate! As we saw, the interaction cross sections of neutralinos are of electro-weak strength, therefore large detector masses and long measuring times are required. Given the standard halo parameters, the rate estimates can be cast into a handy formula
\begin{equation}
R ~=~ \frac{403}{A}\left( \frac{{\rm GeV}}{M_\chi} \right) \left( \frac{\rho_x}{0.3\frac{{\rm GeV}}{{\rm cm}^3}}\right)  \left( \frac{<v_x>}{230\frac{{\rm km}}{{\rm s}}}\right)  \left( \frac{\sigma_A^{SD,SI}}{{\rm pb}}\right)\frac{{\rm counts}}{{\rm kg}\cdot {\rm day}} 
\label{eq:eq32}
\end{equation}

This shows us that in order to reach into the heart of the predicted SUSY cross sections for spin dependent and spin independent interactions i.e. $\sigma_{sd}$ $\approx$ 10$^{-5}$pb and $\sigma_{si}$ $\approx$ 10$^{-9}$ pb we need detectors able to record about one event per tonne per day. Current projects are still at least a factor 100 away in sensitivity. \\

In order to achieve their goals, direct detection experiments must therefore fulfill a couple of requirements. As mentioned, they have to work with very low, keV thresholds. Internal and externally induced backgrounds must be passively or actively reduced. Detectors must be protected from neutrons, since they induce WIMP like recoil events. Especially to eliminate cosmic muon induced neutrons, dark matter experiments have to be located at deep underground sites, e.g. in SNOLAB, at a depth of 2000m, where the cosmic neutron background is less than 0.2 n/ton/y!).\\

If a detector finds a signal, how can we make sure that it was a dark matter particle?  First of all we can exploit the fact that the earth is moving with a speed of 30km/s around the sun in a plane slightly inclined with respect to the galactic plane. As a consequence of that we should encounter a WIMP head-wind in summer and a tailwind in winter and given enough statistics, detectors should be able to observe an annual variation in count rate at the level of 5 to 7\%. If the detection proceeds via a spin-independent interaction, different detectors with different targets should see an $A^2$ dependence in count rate. Similarly, spin dependent interactions can be confirmed by choosing targets with different spins. Also specially constructed detectors might trace the recoil direction itself and reconstruct day-night dependent directional changes due to the rotation of the earth. \\  

Presently more than 23 Dark Matter experiments are active with around 8 experiments taking data. A compilation is shown in table \ref{table1}.\\

\begin{table}
\tbl{Direct dark matter search experiments \label{table1}}
{\begin{tabular}{@{}ccccc@{}}
\hline Experiment & Detector & Status & Location & Collaboration \\ [1ex]
\hline
\\[-1.5ex]
 DAMA/LIBRA & NaI & running & Gran Sasso & Italy, China \\ [1ex]
 ANAIS & NaI & constructing & Canfranc & Spain \\ [1ex]
 KIMS & CsI & R \& D & Korea & Korea \\ [1ex]
 HDMS & Ge & running & Gran Sasso & Germany, Russia \\ [1ex]
 Dama-LXe & LXe & running & Gran Sasso & Italy, China \\ [1ex]
 Zeplin II & LXe & running & Boulby & PT, UK, RU, US \\ [1ex]
 Zeplin III & LXe & installing & Boulby & PT, UK, RU, US \\ [1ex]
 XENON 10 & LXe & commiss & Gran Sasso & DE, IT, PT, US \\ [1ex]
 LUX &  &  &  & US \\ [1ex]
 XMASS & LXe &  & Kamioka & Japan \\ [1ex]
 WARP & LAr & running & Gran Sasso & Italy, US \\ [1ex]
 ArDM & LAr & R \& D & Canfranc & CH, ES, PO \\ [1ex]
 DEAP & LAr & R \& D & SNOLAB & Canada, US \\ [1ex]
 CLEAN & LNe & R \& D & SNOLAB? & US, Canada \\ [1ex]
 Rosebud & Bolom. / Scint. & R \& D & Canfranc & Spain, France \\ [1ex]
 EDELWEISS & Bolometer & running & Frejus & F, GE, RU  \\ [1ex]
 CRESST & Bolometer & running & Gran Sasso & DE, UK, IT, RO \\ [1ex]
 CDMS & Bolometer & running & Soudan & US \\ [1ex]
 SIMPLE & Superheated liquid & running & Rustrel & PT, F, US \\ [1ex]
 PICASSO & Superheated liquid & running + R \& D & SNOLAB & CA, US, CZ \\ [1ex]
 COUPP & Superheated liquid & R \& D & Fermilab & US \\ [1ex]
 Drift & Xe gas & R \& D & Boulty & UK, US \\ [1ex]
 MIMAC & $~^3$He gas & R \& D &  & France \\ [1ex]
\hline 
\end{tabular}}
\end{table}

With respect to the applied detection techniques, the experiments fall roughly into three large classes: Ionisation detectors (Ge-diodes, drift chambers), scintillation detectors (NaI, CsI, LXe, LAr) and heat detectors (cryogenic detectors, superheated liquids). Some of the experiments are "hybrids" and exploit different responses in more than one channel to separate signal from background. It is impossible to do justice to all these very sophisticated approaches in several lines only, so we can only give here a rough sketch.\\  

\begin{figure}[ht]
\centerline{\epsfxsize=4.1in\epsfbox{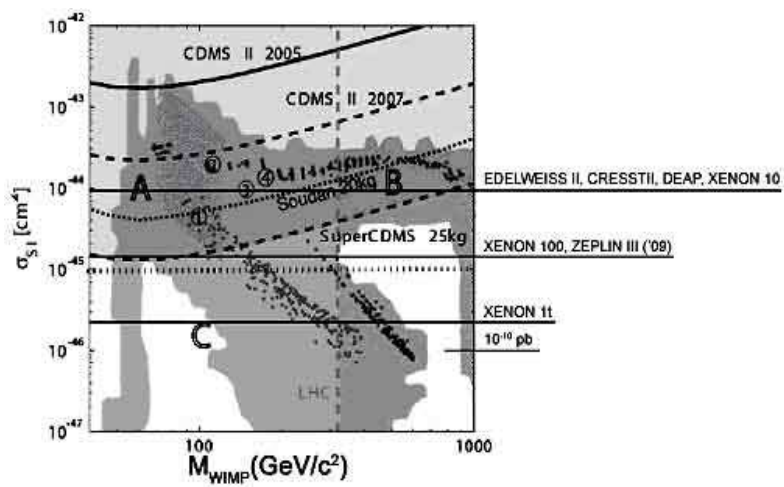}}   
\caption{Summary of existing and projected limits on spin-independent interactions. Broken lines and horizontal line are projected limits\protect\cite{DB}. LHC will be able to push the limits on the neutralino mass above 300 GeV (vertical line).  \label{fig12}}
\end{figure}

\begin{figure}[ht]
\centerline{\epsfxsize=4.1in\epsfbox{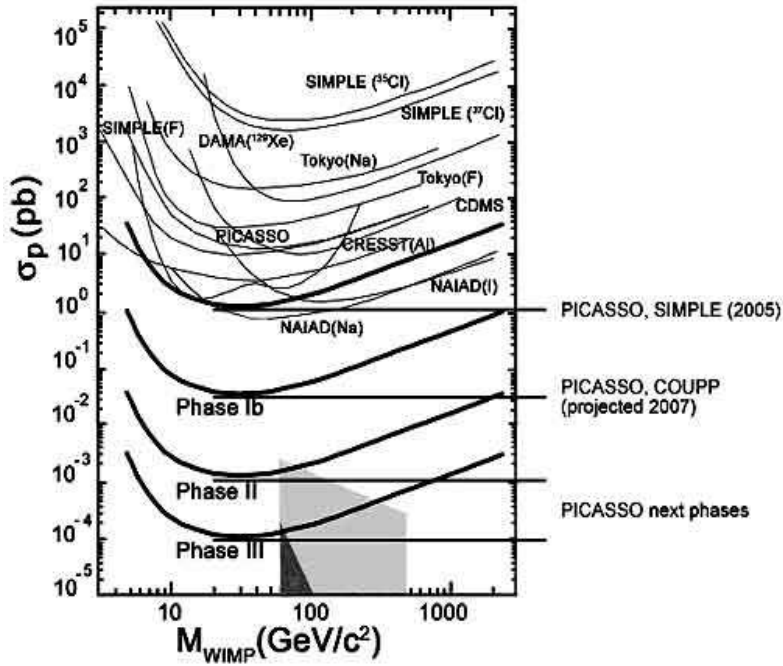}}   
\caption{Summary of existing and projected spin-dependent limits. PICASSO phase 1b, II and III are related to the ongoing and next stages with 3 kg, 25kg and 100 kg of active mass of C$_4$F$_{10}$.  \label{fig13}}
\end{figure}

\underline{Scintillator experiments}: Ionizing radiation interacting with crystals like NaI, CsI, CaF$_2$(Eu) or noble liquids like Xe, Ar, Ne induces the emission of scintillation photons mostly in the UV range.  These photons can be detected either directly or, after being shifted in wavelength, by photomultiplier tubes with quantum efficiencies of around 15\% or with semiconductor photodiodes with quantum efficiencies close to 90\%. Typical light gains reach 2-8 photoelectrons per incident photon. Nuclear recoil events develop usually light pulses with shorter decay times than electron or gamma induced events. Therefore a decay time analysis allows for background discrimination on a statistical basis. Experiments of this kind are DAMA, NAIAD, ANAIS, KIMS, DEAP. Some of the limits obtained are shown in fig. \ref{fig12} and \ref{fig13}.\\

In this category the results of the DAMA experiment are certainly the most discussed ones! The DAMA collaboration deployed 100 kg of NaI crystals in the Gran Sasso National Laboratory, a system of huge caverns accessible by a road tunnel in the Italian Abruzzi mountains. With both target materials, Na and I, this experiment is sensitive in the spin dependent and spin independent sector. Data were taken from 1995 to 2002, accumulating a total of 107731 kgd of exposure, the largest exposure of all direct dark matter experiments so far. The period of data taking covered 7 annual cycles and most interestingly shows in the energy range of 2 - 6 keV a 6.3 $\sigma$ effect modulation in phase with the expected modulation of a WIMP signal due to the earth's motion around the sun.  The fit to the data returns a period of T = 1.00 $\pm$ 0.01 y with an amplitude of A~=~0.0195~$\pm$~0.003~cts~kg$^{-1}$d$^{-1}$keV$^{-1}$. This signal would correspond to a neutralino mass in a region around M$_\chi$ = 52 GeV and a spin-independent cross section of $\sigma_{SI}$ = 7x10$^{-6}$ pb \cite{BER1,BELL}. However this result is disputed by a whole series of experiments in the spin and spin-independent sectors. But there are loophole scenarios: it could still be that the neutralino has a mass below 10 GeV, where other experiments would not have sensitivity and the halo composition might not be what we think it is, maybe with lower velocities and different particle densities\ref{BOT}. Or DAMA sees something else: maybe a dark matter tidal stream from the Sagittarius dwarf galaxy, which is orbiting around the Milky Way\cite{GEL}? Suspense!\\

\underline{Cryogenic experiments} operate crystals of Ge, Si, Al$_2$O$_3$ or TeO$_2$ at temperatures of several mK. Particle interactions create phonons, i.e. heat and the tiny rise in temperature can be measured with superconductor or semiconductor thermometers. For example a 1 keV energy deposition in Ge at 2mK gives us a $\Delta$T of 10$^{-6}$ K. About 1meV is needed to create a phonon and energy resolutions of 4.5eV have been achieved for 6KeV x-rays! Since the ionization or scintillation yields depend on the nature of the incident particle, a comparison of the phonon signal with the ionization or light signal allows a powerful background rejection. Experiments, which exploit this technique, are e.g. CDMS, CRESST, ROSEBUD, CUORE, EDELWEISS and a review of them can be found in ref. \refcite{HK,MYJ}. Some of their limits are shown in fig. \ref{fig12}.\\

For example the CDMS II experiment operates since October 2006 several stacks (called towers!) of 250g modules of Ge and Si crystals at 50mK with a total mass of 5.6 kg. The experiment is installed in the Soudan mine close to Chicago. The ionization signal is compared to the phonon signal and also the signal rise times are used for event discrimination. In this way gamma induced events can be rejected with an amazing efficiency of 99.9998\%, beta decay electrons with 99.75\% efficiency. The experiment is expected to reach a sensitivity of 2x10$^{-8}$ pb for spin-independent interactions during 2007.  The upgrade to SuperCDMS with larger modules and 25 kg total mass should reach a sensitivity of 1x10$^{-9}$ pb by 2012\cite{DB}.\\

\underline{Liquid noble gases} have become another major and promising avenue for large mass experiments. Here LXe, LAr or LNe are the target volume. In the so-called single- phase operation the ionization signal and/or the scintillation photons are collected in the liquid and background reduction is achieved by pulse shape discrimination. In the dual-phase mode, WIMP recoils in e.g. liquid Xe create a scintillation signal and the ionization electrons are drifted in a homogeneous electric field into the gaseous phase above. In a strong 10kV/cm field close to the anode wires the secondary ionization creates a proportionally amplified scintillation light pulse. Nuclear recoils can then be discriminated against $\gamma$-, $\beta$-, $\alpha$- interactions by comparing the relative pulse heights between the primary and secondary light in a delayed coincidence. The number of experiments in this category is literally exploding! As of January 2007 we can list XENON10, ZEPLIN, WARP, XMASS, ArDM, CLEAN, DAMA/LXe, LUX, DEAP. More information on the individual experiments can be found e.g. in ref. \refcite{MYJ,IDM} and some of the limits are displayed in fig. \ref{fig12} and \ref{fig13}.\\

To give an example, let's pick from the list above the DEAP experiment at SNOLAB. It detects scintillation light in LAr at a temperature of 85K. The light yield is 4x10$^4$ photons per deposited MeV and the detector threshold is around 10 keV. During the ionisation process, Ar$_2$* dimers are formed in excited singlet and triplet states, which decay with different lifetimes. Since the fraction of singlet and triplet excitations depends on the ionization density, pulse shape discrimination allows an efficient separation of nuclear recoil and gamma-induced events. A detector of 7 kg is presently installed at SNOLAB and with one year of data taking cross sections of $\sigma_{SI}$ = 1x10$^{-8}$ pb should be reached. A future DEAP 3 detector would accommodate 1 ton of LAr in a spherical 5m-diameter tank, read out by 500 photomultiplier tubes\cite{MBS}\\.

\underline{Superheated liquids} are another maturing detection technique. The principle is based on the traditional bubble chamber operation, but here it is tailored to the detection of WIMP induced nuclear recoils. The detector medium is a metastable liquid, i.e. a liquid heated above its boiling point, and a phase transition is triggered by heat spikes produced by the energy deposited along the track of the traversing particle. More precisely, the degree of metastability or superheat, depends on the difference of the temperature dependent vapor pressure and the applied external pressure. At a given temperature, bubble formation on a particle track occurs, if within a region of critical size l$_c$ the deposited energy, E$_{dep}$, exceeds a threshold energy E$_{min}$
\begin{equation}
E_{dep}~=~\frac{dE}{dx}l_{crit}~\geq ~ E_{min}
\label{eq:eq33}
\end{equation}

In this relation dE/dx is the mean energy deposited per unit distance. Since large specific energy losses are characteristic for nuclear recoils, the operating conditions can be tuned such that the detector is fully sensitive to nuclear recoils, but essentially blind to $\gamma$- or $\beta$- induced events with small dE/dx. For example when operating such a device at a recoil threshold of 5 KeV, $\gamma$-induced events are rejected by more than a factor of 10$^7$! Although this detector is a threshold device, recoil energy spectra can be recorded by ramping the temperature and thus varying the detector threshold.\\

There are two technical realisations:  the COUPP experiment operates since 2005 a 2kg bubble chamber filled with CF$_3$I. As in the usual bubble chamber operation, the detector has to be recompressed after each event. With Fluorine and Iodine as a target, spin-dependent and spin-independent interactions can be explored simultaneously. Events are triggered acoustically and recorded optically\cite{CP}.\\

  The PICASSO and SIMPLE experiments employ the superheated droplet technique, where a fluorine loaded active liquid is dispersed in the form of $\approx$ 50-100$\mu$m diameter droplets in a polymerized or viscous medium.  If an event occurs in a droplet, it will explode and piezolelectric transducers detect the acoustic signal. Apart from occasional recompression periods, this detector is continuously active and can be calibrated easily at high count rates with radioactive sources. Both experiments have so far published nearly identical limits on the spin dependent cross section on protons\cite{SIM,PIC} of $\sigma_{SD}$  $\approx$ 1pb.\\  

PICASSO is presently installing 32 detectors for an active mass of 3 kg of C$_4$F$_{10}$ at SNOLAB. Each detector has a volume of 4.5 l and is read out by 9 piezoelectric transducers. A measurement of the relative time delays allows to locate events inside the detectors with a resolution of around 5 mm. A first group of detectors is installed and data taking is ongoing. At the level of the present intrinsic background ($\alpha$-emitters) a sensitivity of 5x10$^{-2}$ pb is expected for an exposure of 280kgd. Fig. \ref{fig13} gives a summary of present and projected limits in the spin-dependent sector.\\ 

\section{Conclusions}

We have seen that astronomical observations from our galactic backyard to the largest distances our telescopes can explore, predict consistently that a large fraction of the mass of the universe is hidden. Deep field galaxy and large scale redshift surveys combined with gravitational lensing are about to revolutionize our observational techniques and allow us now to reconstruct the spatial distribution of dark matter and even its evolution in time. The precise determination of the cosmological parameters from the study of the cosmic microwave background anisotropy (WMAP) has initiated a new era of precision cosmology. All evidence points to the conclusion, that about 85\% of all gravitationally traceable matter is in form of some non-relativistic non-baryonic, exotic kind of matter, which we call Cold Dark Matter. Particle physics offers several plausible candidates, but its precise nature is at the moment still unknown. The hunt for possible candidates has been opened on three fronts: Direct searches in underground laboratories will be able to explore within the next 7 to 8 years a large part of the supersymmetric cross sections, which are compatible with the cosmological constraints. A new generation of ground and space based indirect search experiments will open new windows to search for the presence of dark matter in our galactic vicinity or to detect WIMP annihilations in the sun's interior.  At LHC we will have the chance to discover dark matter particles in situ and if not, we can at least obtain improved limits on the allowed mass range. We also note the increasing synergy and complimentarily between astronomical observations, direct and indirect searches and experiments at accelerators, which make this field of research to one of the most fascinating in contemporary science. \\    

\section{Acknowledgements}

My most cordial thanks go to Prof. Faqir Khanna for inviting me to such a magnificent place as Lake Louise and for giving me the opportunity to present this series of lectures on Dark Matter in such an inspiring and enjoyable ambiance. I am very grateful to my colleagues Prof. Louis Lessard and Marie-H\'el\`ene Genest for their constructive comments.  Special thanks go to Jonathan Ferland for his competent and skilful assistance in going over the text and for getting it into the right shape. \\

\end{document}